\documentclass[twocolumn,aps,prl,preprintnumbers,bibnotes10pt,superscriptaddress]{revtex4}

\usepackage{amsmath}
\usepackage{graphicx}
\usepackage{ulem}
\usepackage{dcolumn}
\usepackage{epsfig}
\usepackage{bm}
\usepackage{array}
\usepackage[frenchb]{babel}
\usepackage{hyperref}
\hypersetup{
colorlinks=true,
citecolor=black,
linkcolor=red,
urlcolor=black
 , bookmarks=true,
 pdfmenubar=true
}

\usepackage{graphicx}
\usepackage{amsmath}
\usepackage{bm}           
\usepackage{bbm}
\usepackage{ulem}
\usepackage{color}

\newcommand{\bra}[1]{\ensuremath{\langle#1|}}
\newcommand{\ket}[1]{\ensuremath{|#1\rangle}}

\newcommand{\mean}[1]{\ensuremath{\big\langle #1 \big\rangle}}

\newcommand{\vect}[1]{\bm{#1}}

\newcommand{\be}{\begin{equation}}
\newcommand{\ee}{\end{equation}}
\newcommand{\ud}{\mathrm{d}}
\newcommand{\beq}{\begin{eqnarray}}
\newcommand{\eeq}{\end{eqnarray}}

\newcommand{\ii}{{\rm i}}
\newcommand{\ME}{ME }

\begin{document}

\title{Multipartite entanglement in topological quantum phases}

\author{Luca Pezz\`e}
\affiliation{QSTAR and INO-CNR and LENS, Largo Enrico Fermi 2, 50125 Firenze, Italy} 
\author{Marco Gabbrielli}
\affiliation{QSTAR and INO-CNR and LENS, Largo Enrico Fermi 2, 50125 Firenze, Italy} 
\affiliation{Dipartimento di Fisica e Astronomia, Universit\`a degli Studi di Firenze, via Sansone 1, I-50019, Sesto Fiorentino, Italy} 
\author{Luca Lepori}
\affiliation{Dipartimento di Scienze Fisiche e Chimiche, Universit\`a dell'Aquila, via Vetoio 42, I-67010 Coppito-L'Aquila, Italy}
\affiliation{INFN, Laboratori Nazionali del Gran Sasso, Via G. Acitelli 22, I-67100 Assergi (AQ), Italy}
\author{Augusto Smerzi}
\affiliation{QSTAR and INO-CNR and LENS, Largo Enrico Fermi 2, 50125 Firenze, Italy} 

\begin{abstract}
We witness multipartite entanglement in the Kitaev chain -- a benchmark model of one dimensional topological insulator -- also with variable-range pairing,
using the quantum Fisher information. 
Phases having a finite winding number, both  for short- and long-range pairing, are 
characterized by a power-law diverging finite-size scaling  of multipartite entanglement. 
Moreover the occurring quantum phase transitions are sharply marked by the divergence of the 
derivative of the quantum Fisher information, even in the absence of a  closing energy gap. 
\end{abstract}

\maketitle
\date{\today}

{\it Introduction.--} The characterization of quantum phases and quantum phase transitions (QPTs) 
via entanglement measures and witnesses \cite{HorodeckiRMP2009, GuhnePR2009} is 
an intriguing problem at the verge of quantum information and many-body physics \cite{AmicoRMP2008, Zeng}.
The study of entanglement pushes our understanding 
of QPTs \cite{OsbornePRA2002, OsterlohNATURE2001} 
beyond standard approaches in statistical mechanics \cite{Mussardo, Sachdev} and 
sheds new light on the creation, manipulation and protection of useful resources for quantum technologies.

The literature \cite{AmicoRMP2008} has mostly focused on bipartite entanglement. 
A benchmark is the so-called area law \cite{EisertRMP2010} that relates the amount of
bipartite entanglement among two partition of a many-body system to the surface area between the two blocks \cite{VidalPRL2003, LatorreJPA2009}.
For models with short-range interaction in one dimension, 
the Von Neumann block entropy is constant in the gapped phases while it increases
logarithmically with the system size at criticality \cite{CalabreseJSM2004}. 
Yet, violations of the area law are not always related to a closing gap: a logarithmic increase of the Von Neumann 
entropy occurs also in some gapped phases
of one-dimensional models with long-range interaction \cite{VodolaPRL2014, VodolaNJP2016, KoffelPRL2012, AresPRA2015},
as well as for peculiar short-range models \cite{MovassaghPNAS2016}.
An alternative approach to bipartite entanglement is the analysis of the two-body reduced density matrix 
\cite{OsterlohNATURE2001, WuPRL2004}, often also indicated as pairwise entanglement.

The complex structure of a many-body quantum state is of course much richer than that caught by bipartite/pairwise entanglement. 
Yet, multipartite entanglement (ME) has been much less studied \cite{GuhneNJP2005, HofmannPRB2014}. 
Recently \cite{HaukeNATPHYS2016, LiuJPA2013, MaPRA2009}, ME in models exhibiting 
Ginzburg-Landau-type QPTs has been witnessed using 
the quantum Fisher information (QFI) calculated for the {\it local} order parameter, exploiting  
well-known relations between the QFI and \ME \cite{PezzePRL2009, HyllusPRA2012, TothPRA2012}. 
In this case, \ME diverges at criticality in benchmark spin systems,
as the Ising \cite{HaukeNATPHYS2016}, XY \cite{LiuJPA2013} and the Lipkin-Meshkov-Glick \cite{MaPRA2009, HaukeNATPHYS2016} model. 
However, it has been emphasized that this approach, based on local operators, 
fails to detect \ME at topological QPTs \cite{HaukeNATPHYS2016}.
For this task, methods based on the QFI generally require the extension of entanglement criteria for
{\it nonlocal} operators, along the lines of Ref.~\cite{PezzePNAS2016}.

Here we focus on a paradigmatic model showing topological quantum phases:  
the Kitaev chain of spinless fermions in a lattice \cite{Kitaev,AliceaRPP2012} with 
variable-range pairing \cite{VodolaPRL2014, VodolaNJP2016}.
We calculate the QFI of the ground state for a suitable choice of nonlocal observables and witness ME (the parties being the single sites of the chain) when 
the corresponding correlation functions have a sufficiently slow decay.
We show that phases identified by a nonzero winding number are 
characterized by a superextensive scaling of the QFI, signaling the divergence of ME with the system size. 
This divergence is found, for short-range pairing, only at topologically nontrivial phases, hosting massless edge modes. 
Instead, for long-range pairing superextensivity of the QFI is found whenever the winding number assumes semi-integer values.
Furthermore, for arbitrary pairing range, ME is shown to vary suddenly at QPTs, 
even when the energy gap in the quasiparticle spectrum does not close at the boundary lines.
Our work addresses an open problem in the literature -- the detection of \ME 
in topological phases and QPTs -- and paves the way towards a characterization of strongly-correlated systems in terms of ME.

\begin{figure}[t!]
\includegraphics[clip,scale=0.45]{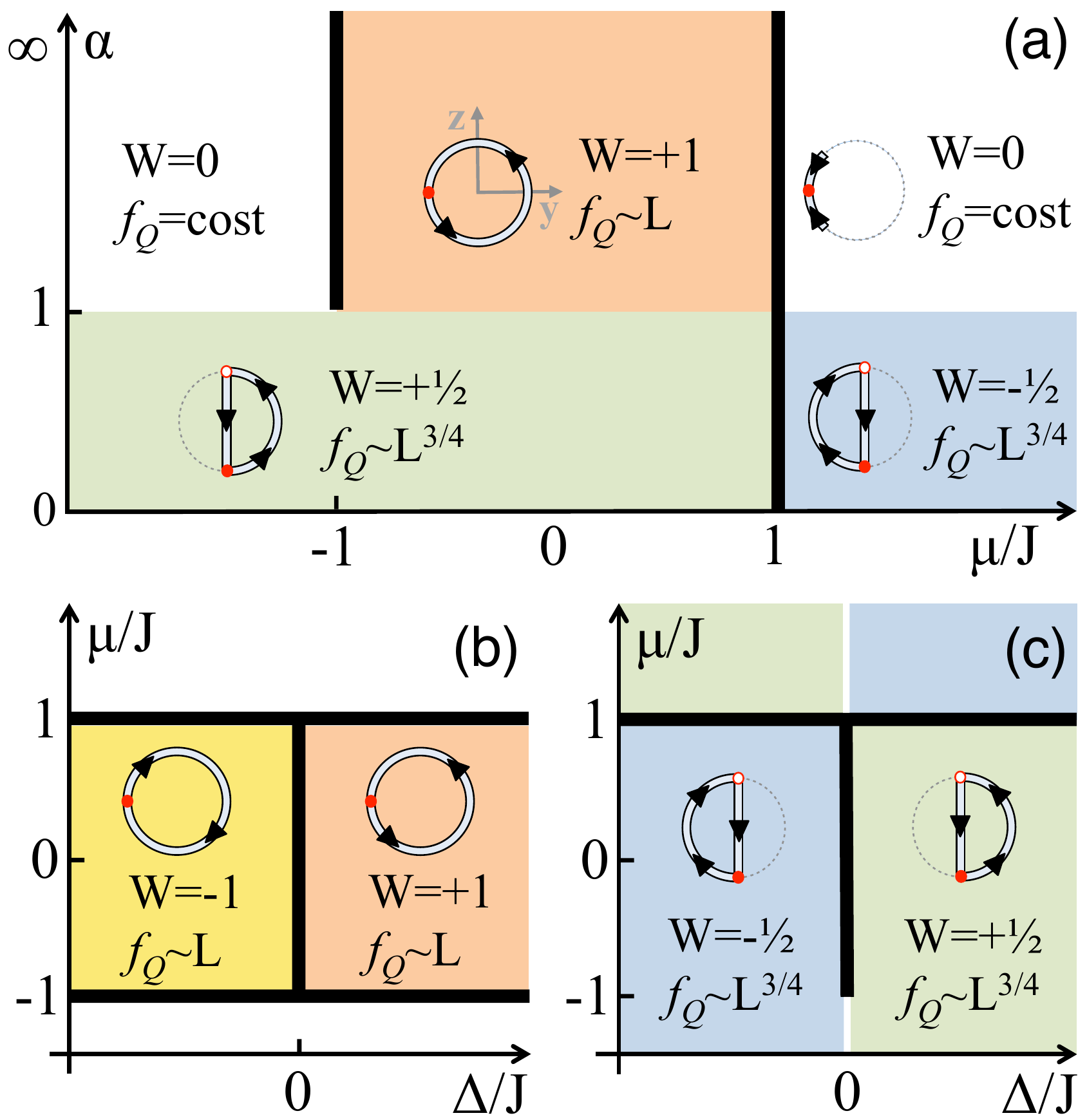}
\caption{Phase diagram of the Kitaev chain
in the $\mu/J$--$\alpha$ plane for $\Delta>0$ \cite{footnote3} (a),
and in the $\Delta/J$--$\mu/J$ plane for (b) nearest-neighbor ($\alpha=+\infty$) and 
(c) infinite-range ($\alpha=0$) pairing.
The thick lines mark a closing gap in the quasiparticle spectrum in the thermodynamic limit $L \to \infty$.
Colored regions highlight different phases with indicated winding number $W$ and scaling of the Fisher density $f_Q$ with the system size $L$.
Thick curved lines shows trajectories in the unit circle in the $y-z$ plane (dotted line) 
as $k$ varies from $0$ (red dot) to $2\pi$ (red circle), see text and \cite{footnote1}.}
\label{Fig1}
\end{figure}

{\it The model.--} The Kitaev chain is a tight-binding model 
with both tunneling and superconducting pairing. 
This was originally studied for nearest-neighbor pairing \cite{Kitaev} and extended 
to variable-range pairing in \cite{VodolaPRL2014, VodolaNJP2016}.
The corresponding Hamiltonian is
\begin{align}
\label{Ham}
\begin{split}
\hat{H} & = - \frac{J}{2} \sum_{j=1}^{L} \left(\hat{a}^\dagger_j \hat{a}_{j+1} + \mathrm{h.c.}\right)  - \mu \sum_{j=1}^L \left(\hat{n}_j - \frac{1}{2}\right)  \\ 
& \,\,\,\,\,\, +\frac{\Delta}{2} \sum_{j=1}^L \,\sum_{\ell=1}^{L-j} d_\ell^{-\alpha} \left( \hat{a}_j \hat{a}_{j+\ell} + \hat{a}^\dagger_{j+\ell} \hat{a}^\dagger_{j}\right), 
\end{split}
\end{align}
where $J>0$, $\hat{a}_j$ is a fermionic annihilation operator on the site $j$, $\hat{n}_j = \hat{a}^\dag_j \hat{a}_j$ and  $L$, 
assumed even, is the total number of sites.
We consider a closed chain, with $d(l) =l$ for $1 \leq l \leq L/2$ and $d(l) =L-l$ for $L/2 \leq l < L$, and 
assume antiperiodic boundary conditions $\hat{a}_{j+L} = - \hat{a}_j$.
Following \cite{Lieb1961}, the Hamiltonian (\ref{Ham}) can be diagonalized exactly by a Bogoliubov transformation. 
The resulting quasiparticle spectrum is \cite{VodolaPRL2014}
\be \label{spect}
\epsilon_{k} = \sqrt{ \big(J \cos k + \mu \big)^2 + \big(f_\alpha(k) \, \Delta/2\big)^2},
\ee
where $k = \frac{2\pi}{L} (n+\frac{1}{2})$ with $ n=0,1,...,L-1$, and $f_\alpha(k) = \sum_{l=1}^{L-1} \frac{\sin(k l)}{d(l)^\alpha}$.
The function $f_\alpha(k)$ displays singularities at $k = 0$
 and $k=2\pi$ in the thermodynamic limit if $\alpha \leq 1$.

The schematic phase diagram of the model is shown in Fig.~\ref{Fig1}.
Colored regions refer to phases with different (and constant) values of the winding number
$W= \int_{0}^{2\pi} \tfrac{d\Theta(k)}{d k} \frac{dk}{2\pi}$, where $\tan \Theta(k) = (\Delta /2)f_\alpha(k)/(J \cos k + \mu)$ is defined as follows \cite{AliceaRPP2012, supp}.
After the Fourier transform $\hat{a}_k^\dag = \tfrac{1}{\sqrt{L}} \sum_j e^{-i k j} \hat{a}_j$, 
Eq. \eqref{Ham} becomes $\hat{H}= \sum_k (\hat{a}_k^\dag, \hat{a}_{-k}) \hat{\mathcal{H}}_k \binom{\hat{a}_k}{\hat{a}_{-k}^\dag}$
where $\hat{\mathcal{H}}(k) = \vect{h} (k) \cdot \vect{\sigma}$ and $\vect{\sigma}$ is the Pauli vector.
We can then associate to the quasi-momentum $k$ the vector $\vect{h}(k) =  \vect{y} \sin \Theta(k)+  \vect{z} \cos \Theta(k)$.
Varying $k$ from 0 to $2\pi$, $\vect{h}(k)$ winds $W$ times around a unit circle in the $y-z$ plane, see Fig.~\ref{Fig1}.
The value of the topological invariant $W$ changes when the system crosses the boundary between two phases.

\begin{figure}[t!]
\includegraphics[clip,scale=0.45]{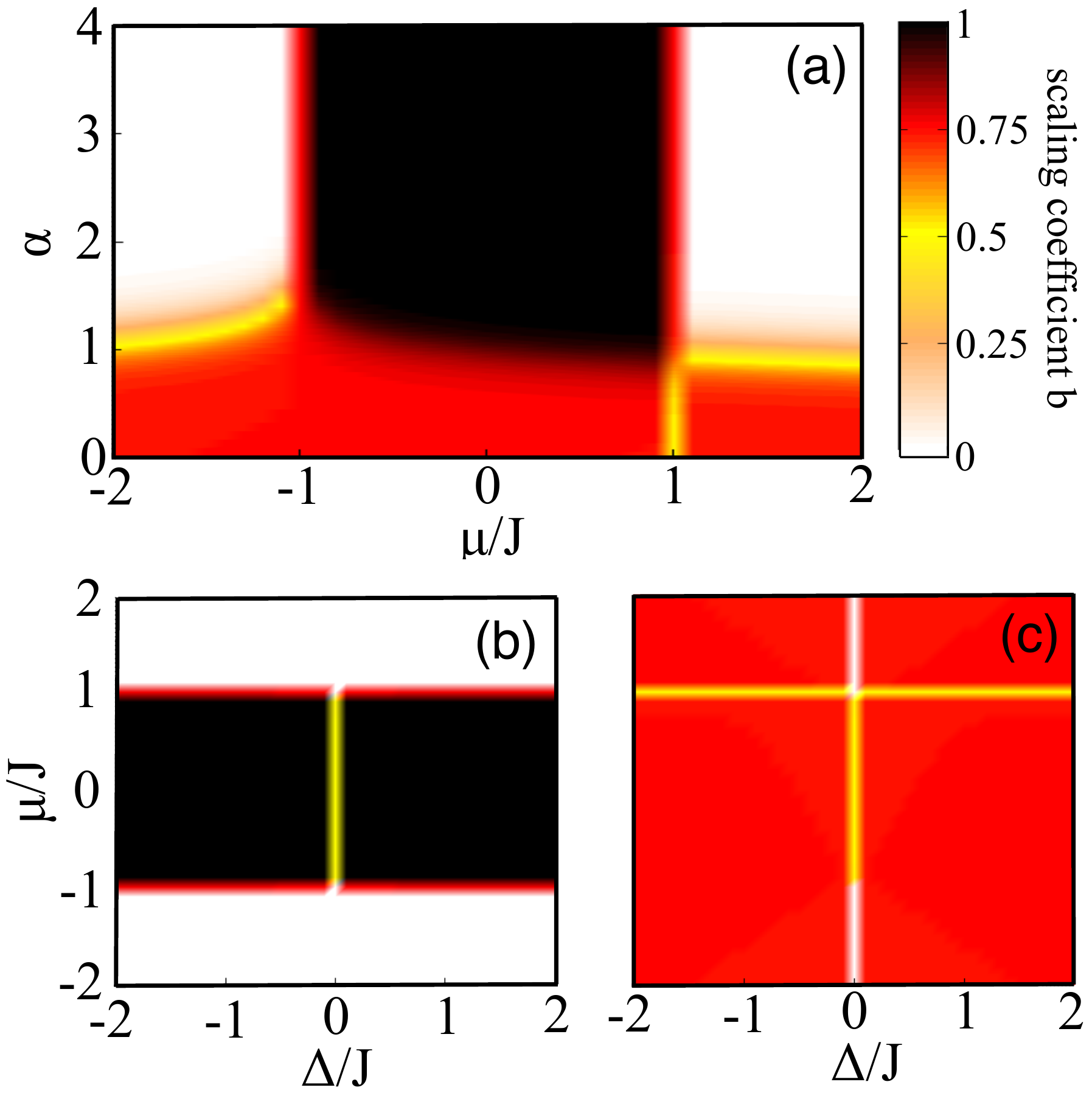} \\
\caption{Phase diagram of the Kitaev chain obtained numerically from the scaling of 
the Fisher density as a function of the system size $L$, $f_Q = 1 + c L^b$, Eq.~(\ref{FQineq}).
The color scale shows the scaling coefficient $b$ in the $\mu/J$--$\alpha$ plane for $\Delta = J$ (a) and 
in the $\Delta/J$--$\mu/J$ plane for nearest-neighbor (b) and 
infinite-range (c) pairing.}
\label{Fig2}
\end{figure}

For short-range pairing ($\alpha>1$), $W$  assumes only integer values: $W=0, \, \pm 1$.
$W\neq 0$ signals a topologically nontrivial phase \cite{AliceaRPP2012}, 
characterized by the presence of massless (Majorana) edge modes in the open chain  \cite{Kitaev}. 
For long-range pairing ($\alpha\leq 1$), semi-integer values $W=\pm 1/2$ appear \cite{delgado2015,lepori2016,footnote1}. 
We quote the corresponding phases as ``long-range''. A similar phase for $\mu/J<1$ supports massive edge modes in the open chain,
while for $\mu/J>1$ no edge mode occurs \cite{delgado2015,lepori2016}.
Long-range phases are all characterized by a violation of the area law for the Von Neumann entropy~\cite{VodolaPRL2014, AresPRA2015}.

For $\Delta \neq 0$ and in the limit $L\to \infty$, 
the energy gap between the superfluid ground state $\vert \psi_{\rm gs}\rangle$ and the first excited state closes 
at $\mu/J = 1$ and $k=\pi$ for all the values of $\alpha$, as well as  
at $\mu/J = -1$ and $k=0$ when $\alpha > 1$ only [see Fig.~\ref{Fig1} (a)]. 
There, $W$ changes between 0 and 1.
Furthermore,  QPT are also present along the line $\alpha=1$: between two phases with $W=1/2$ and $0$ for $\mu/J<-1$, 
between $W=1/2$ and $1$ for $|\mu|/J<-1$, and between $W=-1/2$ and $0$ for $\mu/J>1$.
Remarkably, these QPTs occur without closing the energy gap; they are associated  with various discontinuities, 
for instance in the mutual information and in the decay exponents for the two-point correlations
$\langle \hat{a}_{i}^{\dagger} \hat{a}_j \rangle$, $\langle \hat{a}_{i} \hat{a}_j \rangle$ \cite{VodolaPRL2014,lepori2016}.
In the following we will show that, similarly to the QPT at $\mu/J = \pm 1$, the QPT at $\alpha=1$ is signaled by the 
divergence of the derivative of the QFI with respect to $\alpha$, as well as by the divergence of the fidelity susceptibility.

Figures~\ref{Fig1}(b) and (c) show the phase diagram in the $\Delta/J$--$\mu/J$ plane for $\alpha=\infty$ and $\alpha=0$, respectively.
The energy gap closes at $\Delta=0$ for $\vert \mu\vert /J \leq 1$ and where $\cos k = -\mu/J$.
For $\alpha \leq 1$ and $|\mu|/J>1$,
we have a transition -- without closing the energy gap -- between 
$W=\pm 1/2$ \cite{footnote1}, 
crossing $W=0$ on the singular line $\Delta=0$. 

{\it Multipartite entanglement.--} 
We witness ME in the ground state of the Kitaev model (\ref{Ham}) using the QFI,
\be 
\label{QFIps}
F_Q[\ket{\psi_{\rm gs}}, \hat{O}_{\rho}] = \bra{\psi_{\rm gs}} \hat{O}_{\rho}^2 \ket{\psi_{\rm gs}} - \bra{\psi_{\rm gs}} \hat{O}_{\rho} \ket{\psi_{\rm gs}}^2,
\ee
calculated with respect to the nonlocal operator $\hat{O}_{\rho} = \sum_{j=1}^L \hat{o}_{\rho}^{(j)}$ ($\rho = x,y$), where 
\be \label{O}
\hat{o}_{\rho}^{(j)}= 
(-{\rm i})^{\delta_{\rho, y}}  \big( \hat{a}_j^\dag e^{i \pi \sum_{l=1}^{j-1} \hat{n}_l} + 
(-1)^{\delta_{\rho, y}} 
e^{-i \pi \sum_{l=1}^{j-1} \hat{n}_l} \hat{a}_j \big),
\ee
$\delta_{\rho, y} = 1$ for $\rho=y$ and  $\delta_{\rho, y} = 0$ otherwise.
We also calculate the QFI 
with respect to the staggered operator $\hat{O}_{\rho}^{({\rm st})} = \sum_{j=1}^L (-1)^j \hat{o}_{\rho}^{(j)}$: $F_Q[\ket{\psi_{\rm gs}}, \hat{O}_{\rho}^{(\rm st)}]$.
Noticing that $\bra{\psi_{\rm gs}} \hat{O}_{\rho} \ket{\psi_{\rm gs}}=\bra{\psi_{\rm gs}} \hat{O}_{\rho}^{({\rm st})} \ket{\psi_{\rm gs}}= 0$, 
we can rewrite the Fisher density $f_Q[\ket{\psi_{\rm gs}}, \hat{O}_{\rho}] \equiv F_Q[\ket{\psi_{\rm gs}}, \hat{O}_{\rho}]/L$ for the closed chain as
\be \label{QFIcorr}
f_Q[\ket{\psi_{\rm gs}}, \hat{O}_{\rho}] =  1 + \sum_{l=1}^{L-1}  C_{\rho}(l) \, ,
\ee
where $C_{\rho}(l)=\langle \psi_{\rm gs} \vert \hat{o}_\rho^{(1)} \hat{o}_\rho^{(1+l)} \vert \psi_{\rm gs} \rangle$, 
and analogous expressions for $\hat{O}_{\rho}^{(\rm st)}$, with 
$C_{\rho}^{(\rm st)}(l)=\mean{\psi_{\rm gs} \vert  (-1)^l\hat{o}_\rho^{(1)} \hat{o}_\rho^{(1+l)} \vert \psi_{\rm gs}}$. 
Equation (\ref{QFIcorr}) directly links the QFI to the connected correlation function for the operator $\hat{O}_{\rho}$.

The relation between ME and Eq.~(\ref{QFIps}) is obtained by exploiting the 
convexity properties of the QFI \cite{PezzePRL2009, HyllusPRA2012, TothPRA2012},
that holds even when the QFI is calculated with respect to a nonlocal operator \cite{PezzePNAS2016}.
Specifically, the violation of the inequality 
\be \label{FQineq}
f_Q[\ket{\psi_{\rm gs}}, \hat{O}_{\rho}] \leq \kappa, 
\ee
signals $(\kappa+1)-$partite entanglement ($1\leq \kappa\leq L-1$) between sites of the fermionic chain.
In particular, separable pure states $\ket{\psi_{\rm sep}} = \bigotimes_{j=1}^L \ket{n_j}$, 
where $n_j$ is the occupation number of the $j$-th site (that can assume only values $0$ or $1$), 
satisfy $f_Q\big[\ket{\psi_{\rm sep}} , \hat{O}_{\rho} \big] \leq 1$.
Moreover, states $\vert \psi \rangle$
with $L-1 < f_Q[\vert \psi \rangle, \hat{O}_{\rho}] \leq L$ are genuinely $L$-partite entangled \cite{footnote2}, 
$f_Q[\vert \psi \rangle, \hat{O}_{\rho}] = L$ being the ultimate (Heisenberg) bound.

\begin{figure*}[t!]
\includegraphics[clip,scale=0.53]{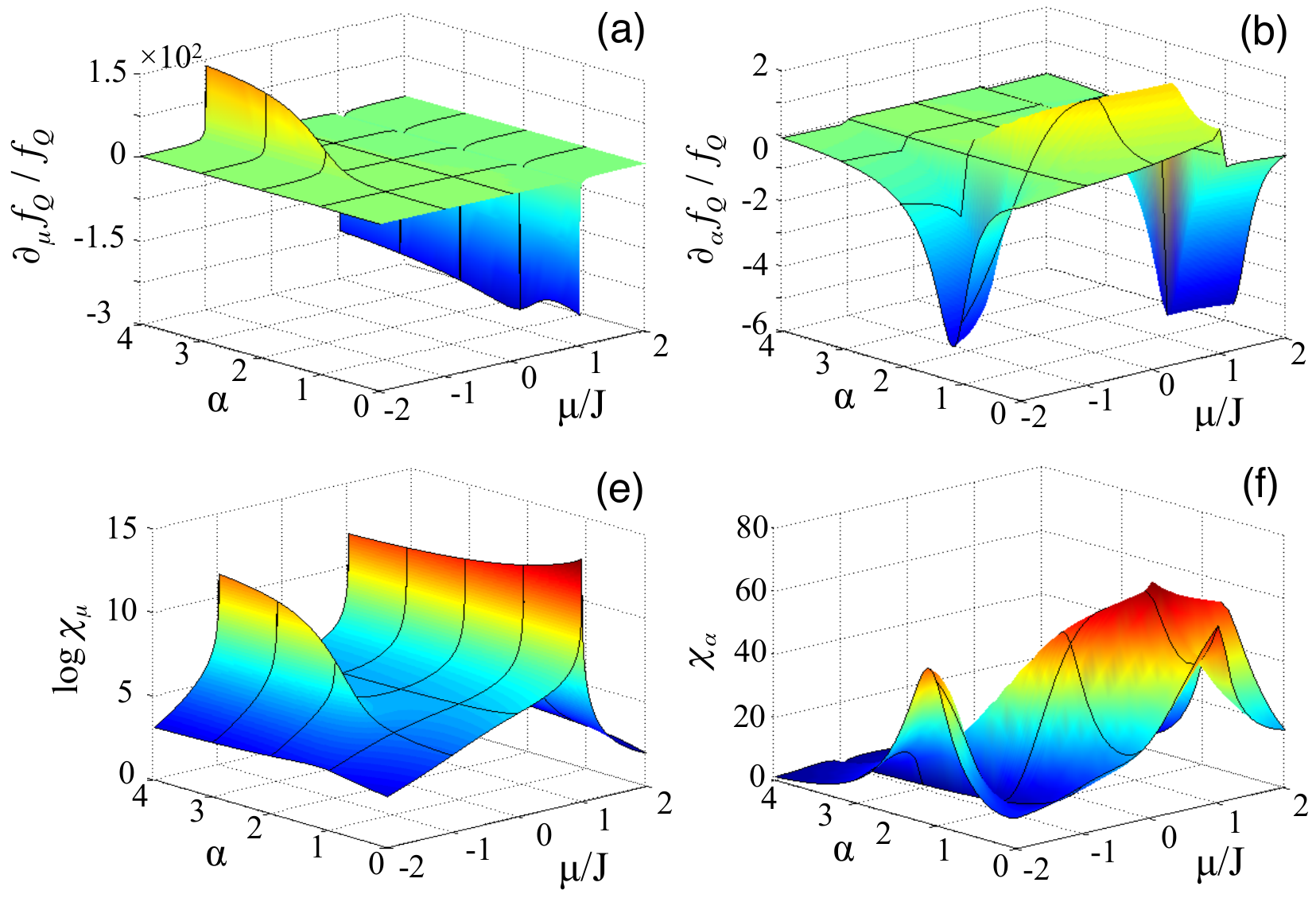} 
\includegraphics[clip,scale=0.53]{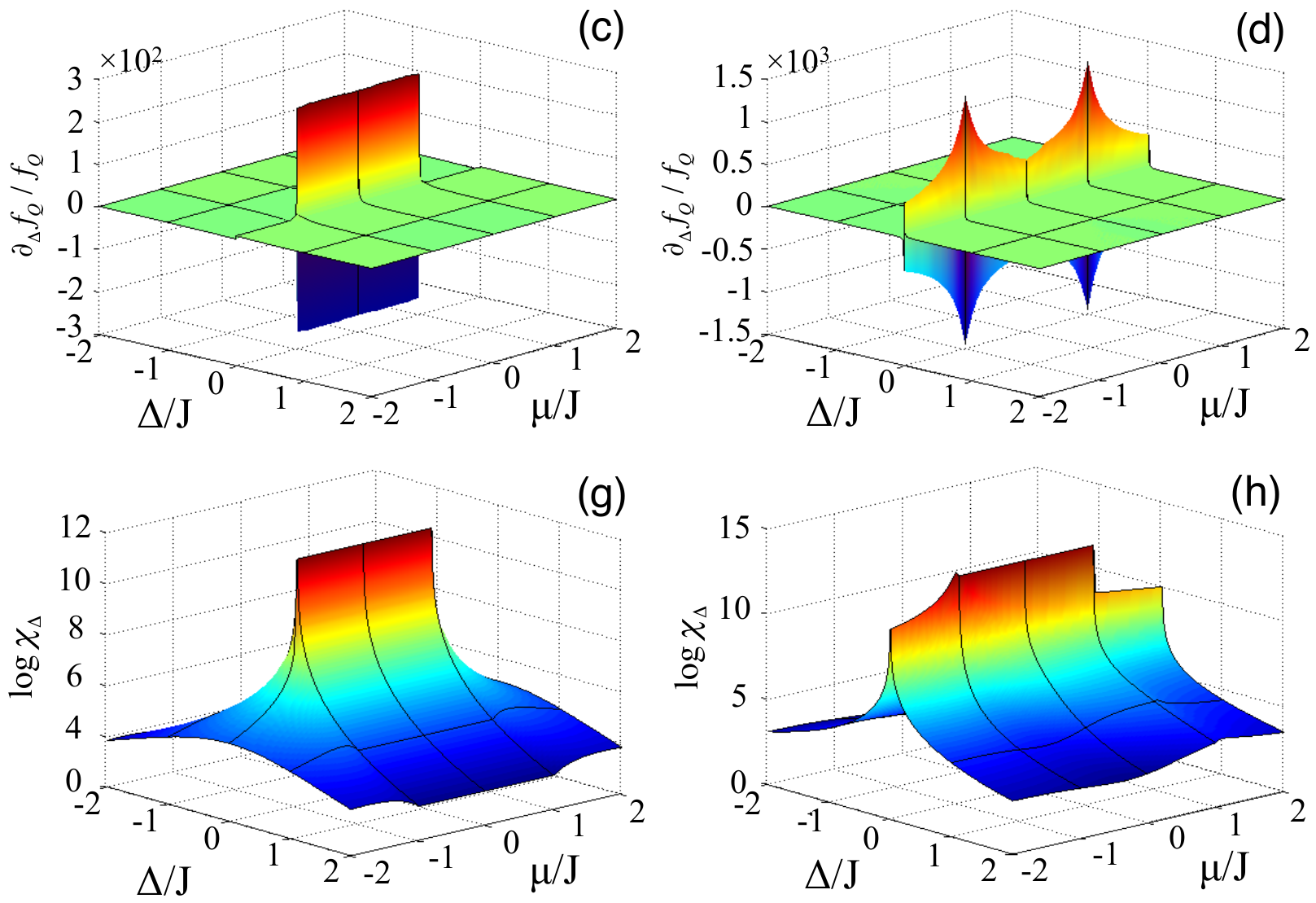}
\caption{Upper panels: weighted derivative of the Fisher density, $\tfrac{1}{f_Q} \tfrac{\partial f_Q}{d \eta}$, with respect to $\eta=\mu/J$ (a), 
$\alpha$ (b) and $\Delta/J$ (c) and (d).
Lower panels: Fidelity susceptibility $\log \chi_{\mu}$ (e), $\chi_{\alpha}$ (f) and $\log \chi_{\Delta}$ (g) and (h).
In all panels, black lines are cuts at integer values of the parameters.
All plots have been obtained for $L=1000$ sites. 
In panels (a,e) and (b,f) $\Delta=J$; 
in panels (c,g) $\alpha=1000$, while in (d,h) $\alpha=0$.
The singularities at $\alpha=1$ for $\tfrac{1}{f_Q} \tfrac{\partial f_Q}{d \alpha}$ [panel (b)] and $\chi_\alpha$ [panel (f)] develop 
slowly in the system size, see \cite{supp} for a finite-size scaling analysis.
}
\label{Fig3}
\end{figure*}

{\it QFI phase diagram.--}
We have calculated the variance Eq.~(\ref{QFIps}) for different values of the parameter of Kiteav chain,
using standard exact techniques \cite{Lieb1961}.
The QFI follows an asymptotic power law scaling 
\be \label{FQ}
f_Q[\ket{\psi_{\rm gs}}, \hat{O}_{\rho}] = 1 + c \, L^b \, ,
\ee
where the coefficients $b$ and $c$ depend on $\mu/J$, $\delta/J$, $\alpha$ and $\rho$, but are independent on $L$. 
The scaling behavior is reported schematically in Fig.~\ref{Fig1}, while Fig.~\ref{Fig2}
shows numerical results obtained up to $L \approx 1000$.
 
The scaling coefficient $b$ is directly related to the behavior of the correlation function.
For instance, an exponentially decaying correlation, $C_{\rho}(l)=e^{-d(l)/\xi}$ on a ring, with $\xi>0$ independent on $L$, 
gives $b=0$ and $c=2/(e^{1/\xi}-1)$ in the thermodynamic limit.
In this case, the QFI is extensive: $c>\kappa-1$ is obtained for 
$\xi^{-1} <\log[(\kappa+1)/(\kappa-1)]$ and witnesses $\kappa$-partite entanglement
that remains constant when increasing the system size.  
Instead, when $b>0$, the QFI is superextensive: the larger is $L$, the larger is the witnessed $\kappa$-partite entanglement.
Values $0<b<1$ can be related to a rescaling of the correlation functions with the system size, 
$C_{\rho}(l)=L^{b-1} c_{\rho}(l/L)$, giving $c = \int_{0}^1 d x \, c_{\rho}(x)$ \cite{supp}.
This is obtained, for instance, when $C(l) \sim 1/l^{1-b}$.

In Figs. \ref{Fig1} and \ref{Fig2} we use the scaling coefficient $b$ to characterize the different phases of the Kitaev model. 
In particular, in Figs.~\ref{Fig1}(a) and~\ref{Fig2}(a) we plot the phase diagram 
of $b$ in the $\mu/J$-$\alpha$ plane, for the operator $\hat{O}_\rho$ maximizing the QFI (see below).
For short-range pairing ($\alpha > 1$), we find $b=1$ for $|\mu|/J<1$ and $b=0$ for $|\mu|/J>1$. 
Therefore, a superextensive QFI is observed only in the topologically nontrivial phase. 
On the critical lines $|\mu| / J = 1$, we have $b=3/4$, that is 
associated to the algebraic asymptotic decay of the correlation functions, 
$C_x(l \to L) \sim L^{1/4}$ \cite{supp}, implied at $\alpha >2$ by conformal invariance \cite{Mussardo,lepori2016ann}.
For $\alpha \approx 1$ the numerical calculations are affected by finite-size effects \cite{supp}. 
Yet, when $\alpha=1$, we can argue (see \cite{supp} for a finite-size scaling analysis) that $b=1/2$ for $|\mu|/J >1$, and $b=3/4$ for $|\mu|/J \leq 1$.
For long-range pairing ($\alpha<1$), $b=3/4$ for $\mu/J\neq 1$ and $b=1/2$ for $\mu/J= 1$.
The above scalings refer for every $\alpha$ to the QFI relative to an optimal choice of operators, 
that is $\hat{O}_x$ for $\mu/J\leq 1$ and $\hat{O}_y^{(\rm st)}$ for $\mu/J\geq1$.

In Figs. \ref{Fig1} and \ref{Fig2} we also plot $b$ in the $\Delta/J$--$\mu/J$ plane 
for nearest-neighbor [$\alpha=\infty$, panel (b)] and infinite-range [$\alpha=0$, panel (c)] pairings.
Notice that the regime $\Delta<0$
is mapped on the one at $\Delta > 0$ by the phase re-definition $\hat{a}_j \to \pm i \, \hat{a}_j $; 
this operation also interchanges $f_Q[\vert \psi_{\rm gs} \rangle, \hat{O}_x] \leftrightarrow f_Q[\vert \psi_{\rm gs} \rangle, \hat{O}_y] $ and 
$f_Q[\vert \psi_{\rm gs} \rangle, \hat{O}_x^{(\rm st)}] \leftrightarrow f_Q[\vert \psi_{\rm gs} \rangle, \hat{O}_y^{(\rm st)}] $.
For $\alpha=\infty$ and $\Delta\neq 0$, we have $b=1$ for $|\mu|/J <1$, $b=3/4$ for $|\mu|/J =1$, and $b=0$ elsewhere.
For $|\Delta|/J=0$ and $|\mu|/J <1$ we have $b=1/2$. 
Again, a superextensive scaling of the QFI is found in the correspondence of phases with nontrivial winding numbers. 
If $\alpha=\infty$ the Kitaev Hamiltonian maps, via the Jordan-Wigner transformation \cite{Lieb1961}, 
to the short-range Ising model (for $\Delta=J$) and our scaling of the QFI agrees with existing calculations \cite{HaukeNATPHYS2016, LiuJPA2013}.
In particular, the correlation function of the $\hat{O}_x$ operator is exponentially decaying for $|\mu|/J >1$ and 
constant for $|\mu|/J <1$: genuine $L$-partite entanglement is witnessed at $\mu=0$ where $C_x(l)=1$ $\forall l$ \cite{supp}.
For $\Delta=0$ (and every $\alpha$), the Kitaev chain maps to the XX model. 
It is known \cite{BarouchPRA1971} that the ground state is a product state for $|\mu|/J>1$ and, accordingly, we find 
$c=0$ in this case \cite{supp}.
In the full diagram of Fig.\ref{Fig2}(b), the optimal choice of operators is $\hat{O}_x$ for $\Delta/J\geq 0$ and $\hat{O}_y$ for $\Delta/J\leq 0$.

For infinite-range pairing ($\alpha=0$) we find $b=3/4$ everywhere, 
except at the phase boundaries: $b=1/2$ for $\mu/J=1$ and $|\Delta|/J>1$, as well as for  $|\Delta|/J=0$ and $|\mu|/J <1$, 
while $b=0$ for $|\Delta|/J=0$ and $|\mu|/J > 1$.
We can distinguish four regions in the phase diagram of Fig.~\ref{Fig2}(c), singled out by the operators optimizing the QFI.
For $|\mu|/J < 1$, the optimal operators are $\hat{O}_x$ for $\Delta>0$ and $\hat{O}_y$ for $\Delta<0$,
while, if  $|\mu|/J > 1$, $\hat{O}_{x,{\rm st}}$ for $\Delta>0$ and $\hat{O}_{y,{\rm st}}$ for $\Delta<0$.

Next, we show that the QPT between phases characterized by different winding numbers are signaled by a divergence of the derivative of the QFI
with respect to the control parameter.   
This is illustrated in Fig.~\ref{Fig3}, where we plot the weighted derivative of $f_Q$ with respect to 
$\mu/J$ [panel (a)], $\alpha$ [panel (b)], and $\Delta/J$ [panel (c) and (d)].
These results can be understood by taking the derivative of Eq.~(\ref{FQ}) with respect to a parameter 
$\eta$ of the model (i.e. $\eta =   \mu/J$, $\Delta/J$ or $\alpha$):
\be \label{dFQ}
\frac{1}{f_Q}\frac{\partial f_Q}{\partial \eta} = \frac{c \, L^b}{1+c \, L^b} \times \bigg( \frac{1}{c} \frac{\partial c}{\partial \eta} + \frac{\partial b}{\partial \eta} \,  \log L \bigg).
\ee
In the interesting case $c\neq 0$ and $b\geq 0$, Eq.~(\ref{dFQ}) diverges in the thermodynamic limit $L\to \infty$
either because of a divergence of $\partial_\eta c / c$ or, even when $\partial_\eta c$ is smooth, 
because of $\partial_\eta b \neq 0$.
Figure~\ref{Fig3}, obtained at $L = 1000$, shows that, while $\partial_\mu f_Q/f_Q$ and $\partial_\Delta f_Q/f_Q$ vary sharply at the phase transition points
of Fig.~\ref{Fig1} (see \cite{supp} for a plot of the coefficients $b$ and $c$), $\partial_\alpha f_Q/f_Q$ varies smoothly as a function $\alpha$.
Yet, a finite-size scaling analysis \cite{supp} shows that, in the limit $L\to \infty$ (up to $L=5000$ in our numerics), 
$\partial_\alpha f_Q/f_Q$ tends to peaks at $\alpha=1$. 
Therefore, a fast change of the QFI is able to detect the QPT at $\alpha=1$
-- associated to a change of the winding number -- even if it occurs without closing of the gap in the quasiparticle spectrum.

The different QPTs of Fig.~\ref{Fig1} are also detected by the fidelity susceptibility. In general, this quantity
detects the qualitative changes in the ground state when crossing a QPT and should not be confused with the QFI, Eq. (\ref{QFIps}), studied here. 
The fidelity susceptibility is defined as $\chi_{\eta} = - \frac{d^2 \mathcal{F}}{d \eta^2}$ \cite{YouPRE2007, ZanardiPRL2007}, where 
$ \mathcal{F} = \bra{ \psi_{\rm gs} } \tilde{\psi}_{\rm gs} \rangle = 
\prod_{n=0}^{L/2}   \cos \frac{\theta_{k_n} - \tilde{\theta}_{k_n}}{2}$ is the fidelity between 
$\ket{\psi_{\rm gs}} $ and $\ket{\tilde{\psi}_{\rm gs}}$, which are ground states of the Hamiltonian in Eq. (\ref{Ham})
for different values of the parameter(s).
Analytical expressions of $\chi_{\mu}$, $\chi_\alpha$ and $\chi_\Delta$ are given in the Supplementary Material~\cite{supp}.
Plots of the fidelity susceptibly are shown in the panels (e)-(h) of Fig.~\ref{Fig3}, they can be compared with the panels (a)-(d), showing the 
weighted derivative of the Fisher density. 
Notice that both display similar divergencies at the critical transition points, also on the massive line $\alpha  = 1$.

{\it Conclusions.--}
The quantum Fisher information detects multipartite entanglement in the topological and long-range phases 
(with nonvanishing winding numbers) of the Kitaev chain with variable-range pairing.
A key aspect is the calculation of the quantum Fisher information relative to nonlocal operators showing long-range correlations, 
whereas the quantum Fisher information relative to local operators is unable to detect entanglement in this model, 
as noted in \cite{HaukeNATPHYS2016}.  

Furthermore, we have shown that QPTs are identified by the divergence of the derivative of the quantum Fisher information 
with respect to different control parameters, 
even when the phase transition is not associated to a closing gap in the excitation spectrum. 
A similar behavior is also found for the fidelity susceptibility.
Our results are a step forward in the study of entanglement in topological insulators by providing a clear evidence of 
multipartite entanglement in these systems. 

{\it Acknowledgements --}
We thank S. Ciuchi, S. Paganelli, and D. Vodola for useful discussions.

\newpage
 
\begin{widetext} 
 
{\bf Supplemental Materials} \\
 
{\it Hamiltonian and basic equations.--}
We discuss here the Bogoliubov Hamiltonian and the calculation of the variance giving the QFI. 
We start from Eq.~(1) of the manuscript, 
\be \label{Ham0}
\hat{H}= - \frac{J}{2} \sum_{j=1}^{L} \left(\hat{a}^\dagger_j \hat{a}_{j+1} + \mathrm{h.c.}\right)  - \mu \sum_{j=1}^L \left(\hat{n}_j - \frac{1}{2}\right)  
+\frac{\Delta}{2} \sum_{j=1}^L \,\sum_{\ell=1}^{L-j} d_\ell^{-\alpha} \left( \hat{a}_j \hat{a}_{j+\ell} + \hat{a}^\dagger_{j+\ell} \hat{a}^\dagger_{j}\right).
\ee 
The parameters $\mu$, $J$ and $\Delta$ are, respectively, the chemical potential, the hopping rate 
(we take $J$ real and positive) and the pairing strength, and $L$, assumed even, is the total number of sites (we consider even values of $L$).
The Fourier transform and its inverse read
\be
\hat{a}_{k_n}^\dag = \frac{1}{\sqrt{L}} \sum_{j=1}^{L} e^{-{\rm i} k_n j} \hat{a}_{j}^\dag, 
\qquad
\qquad
\hat{a}_j^\dag = \frac{1}{\sqrt{L}} \sum_{n=0}^{L-1} e^{{\rm i} k_n j} \hat{a}_{k_n}^\dag, 
\nonumber
\ee
where 
\be
k_n = \frac{2\pi}{L} \bigg( n + \frac{1}{2}\bigg), \quad n=0,1,2,...,L-1.
\ee
We have 
\beq
- \frac{J}{2} \sum_{j=1}^{L} \left(\hat{a}^\dagger_j \hat{a}_{j+1} + \mathrm{h.c.}\right)  - \mu \sum_{j=1}^L \left(\hat{n}_j - \frac{1}{2}\right) 
= - \sum_{k} \bigg( \frac{J}{2} \cos k + \frac{\mu}{2} \bigg) \big( \hat{a}^\dagger_k \hat{a}_k +  \hat{a}^\dagger_{-k} \hat{a}_{-k} \big) + \frac{\mu L}{2},
\eeq
and 
\be 
\sum_{j=1}^L \,\sum_{\ell=1}^{L-j} \frac{\hat{a}_j \hat{a}_{j+\ell} }{d_\ell^{\alpha}}
 = \frac{1}{2} \sum_{j=1}^L \,\sum_{\ell=1}^{L-1} \frac{\hat{a}_j \hat{a}_{j+\ell} }{d_\ell^{\alpha}} 
 = \frac{1}{4} \sum_{j=1}^L \,\sum_{\ell=1}^{L-1} \frac{\hat{a}_j \hat{a}_{j+\ell} -  \hat{a}_{j+\ell}\hat{a}_j }{d_\ell^{\alpha}}
 = \frac{{\rm i}}{2} \sum_k f_{\alpha}(k) \hat{a}_k \hat{a}_{-k}
\ee
where $f_\alpha(k) = \sum_{\ell=1}^{L-1} \frac{\sin k \ell}{d_\ell^{\alpha}}$.
The function $f_\alpha(k)$ diverges at $k=0$ when $\alpha \leq 1$.
In particular, $f_\infty(k)=2 \sin (k)$ and $f_0(k) = \cot (k/2)$.
In the limit $L\to \infty$, we have $f_\infty(k) = 2 \sin k$ and $f_0(k) = \cot (k/2)$.
The Hamiltonian (\ref{Ham0}) becomes 
\be \label{Ham}
\hat{H} = \sum_k \big( \hat{a}_k^\dag, \hat{a}_{-k} \big)
\hat{\mathcal{H}}_k 
\begin{pmatrix}
\hat{a}_k \\
\hat{a}_{-k}^\dag
\end{pmatrix}, 
\ee
where
\be
\hat{\mathcal{H}}_k =  \begin{bmatrix}
-(\tfrac{J}{2}\cos k + \tfrac{\mu}{2}) & {\rm i} \tfrac{\Delta}{4} f_\alpha(k) \\
-{\rm i} \tfrac{\Delta}{4} f_\alpha(k) & (\tfrac{J}{2}\cos k + \tfrac{\mu}{2})\\
\end{bmatrix}
=- \frac{\Delta}{4} f_\alpha(k) \hat{\sigma}_y -\bigg(\frac{J}{2}\cos k + \frac{\mu}{2}\bigg) \hat{\sigma}_z,
\ee
and $\hat{\sigma}_y$ and $\hat{\sigma}_z$ are Pauli matrices. 
We associate this Hamiltonian to a vector
$\vect{h}(k) = h_y(k) \vect{y} + h_z(k) \vect{z}$ in the unit circle, where
\be
 h_y(k) = - \frac{f_\alpha(k) \Delta/2 }{\sqrt{[J \cos k + \mu]^2 + [f_\alpha(k) \Delta/2]^2}} \equiv \sin \Theta(k),
\ee
and 
\be
h_z(k) = - \frac{J \cos k + \mu}{\sqrt{[J \cos k + \mu]^2 + [f_\alpha(k) \Delta/2]^2}} \equiv \cos \Theta(k).
\ee
Finally, the Hamiltonian (\ref{Ham}) can be diagonalized by the Bogoliubov transformation
\be
\begin{pmatrix}
\hat{a}_k \\
\hat{a}_{-k}^\dag
\end{pmatrix} = 
\begin{bmatrix}
\cos \Theta(k)/2 & {\rm i} \sin \Theta(k)/2 \\
 {\rm i} \sin \Theta(k)/2 & \cos \Theta(k)/2 \\
\end{bmatrix}
\begin{pmatrix}
\hat{\eta}_k \\
\hat{\eta}_{-k}^\dag
\end{pmatrix},
\ee
giving
\be
\hat{H} = \sum_k \epsilon_k \bigg(\eta_k^\dag \eta_k -\frac{1}{2}\bigg),
\ee
where 
$\epsilon_k =  \sqrt{[J \cos k + \mu]^2 + [f_\alpha(k) \Delta/2]^2}$.
When $\Delta \neq 0$, the energy gap $\Delta E$ between the ground state (of energy 0) and the first excited state can close (in the limit $L\to \infty$) 
only when $f_\alpha(k)=0$ for some quasi-momentum $k$.
When $\alpha>1$, this happens for $k=0$ (in which case $\Delta E=0$ at $\mu/J=-1$) and $k=\pi$ (in which case $\Delta E=0$ at $\mu/J=1$).
When $\alpha \leq 1$, $f_\alpha(k)$ vanishes only at $k=\pi$ and the energy gap closes only when $\mu/J=1$. \\

{\it $\mu/J$--$\alpha$ phase diagram.--}
We point out that the optimal operator for the QFI is different in the different regions of the phase diagram. 
In Fig.~\ref{FigSupp1} we plot the Fisher density $f_Q[\vert \psi_{\rm gs} \rangle, \hat{O}_x]$ [panel (a)] and 
$f_Q[\vert \psi_{\rm gs} \rangle, \hat{O}_y^{(\rm st)}]$ [panel (b)].
The QFI calculated for the other operators, i.e. $\hat{O}_y$ and $\hat{O}_x^{(\rm st)}$, is always smaller than $L$
and it is not shown.

\begin{figure}[h!]
\includegraphics[clip,scale=0.68]{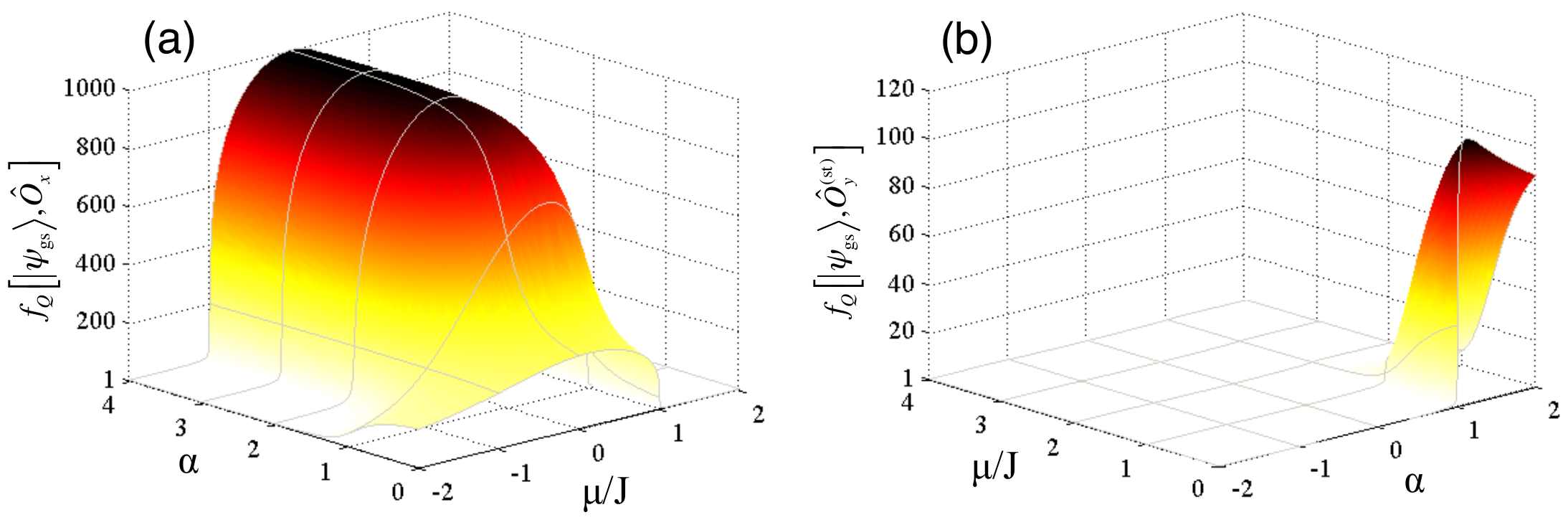}
\caption{(a) $f_Q[\vert \psi_{\rm gs} \rangle, \hat{O}_x]$ and (b)  $f_Q[\vert \psi_{\rm gs} \rangle, \hat{O}_y^{(\rm st)}]$
in the $\mu/J$--$\alpha$ phase diagram, for $L=1000$ and $\Delta=J>0$.
Solid lines are cut at integer values of the parameters.}
\label{FigSupp1}
\end{figure}

We define $f_Q[\vert \psi_{\rm gs} \rangle]$ as the Fisher density $f_Q[\vert \psi_{\rm gs} \rangle, \hat{O}]$ maximized over 
the four operators considered in the manuscript, i.e. $\hat{O}_{\rho}$ and $\hat{O}_{\rho}^{(\rm st)}$, with $\rho=x,y$.
We fit $f_Q[\vert \psi_{\rm gs} \rangle]$ as a function of $L$ as $f_Q[\vert \psi_{\rm gs} \rangle] = 1 + cL^b$ (see below), 
where the coefficients $b$ and $c$ depend on $\mu/J$, $\Delta/J$ and $\alpha$.
In Fig.~\ref{FigSupp2} we plot $b$ and $\log c$ in the phase diagram.
In Fig.~\ref{FigSupp3} and \ref{FigSupp4} we plot the derivative of $b$ and $c$, respectively, with respect to $\mu$ [in panel (a)] and $\alpha$ [in panel (b)].

\begin{figure}[h!]
\includegraphics[clip,scale=0.75]{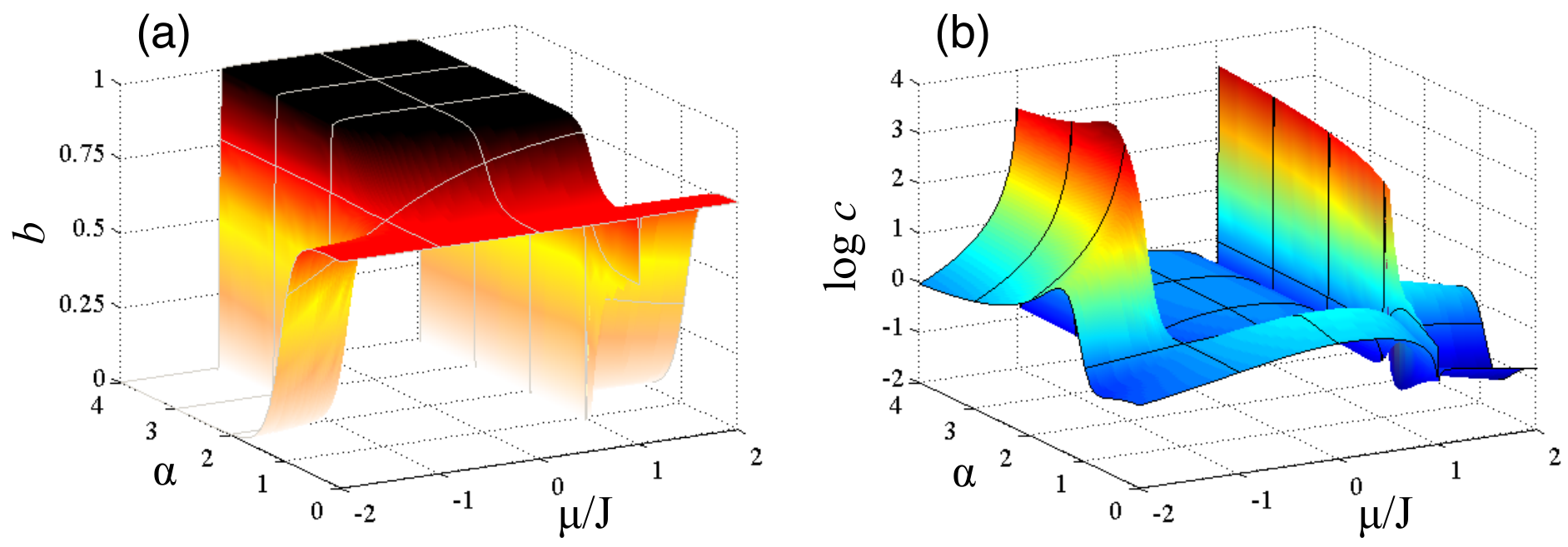}
\caption{Coefficients $b$ and (c) of the polynomial scaling of the Fisher density $f_Q = 1 + cL^b$
in the $\mu/J$--$\alpha$ phase diagram.
They show sharp variations in correspondence of the closing energy gap in the quasiparticle spectrum, $\mu/J=-1$ for $\alpha>1$
and  $\mu/J=+1$ for $\alpha\geq 0$. They change smoothly as a function of $\alpha$ for $\alpha\approx 1$.}
\label{FigSupp2}
\end{figure}

\begin{figure}[h!]
\includegraphics[clip,scale=0.8]{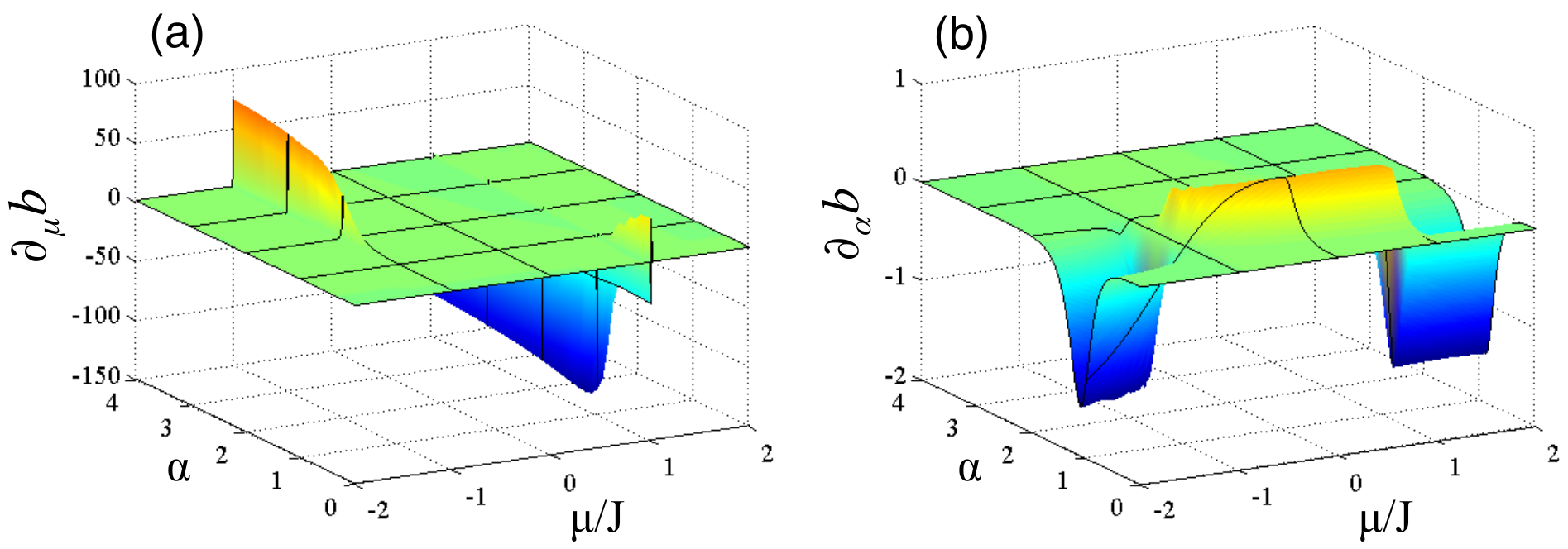}
\caption{Derivative of $b$ with respect to $\mu$ (a) and $\alpha$ (b). }
\label{FigSupp3}
\end{figure}

\begin{figure}[h!]
\includegraphics[clip,scale=0.8]{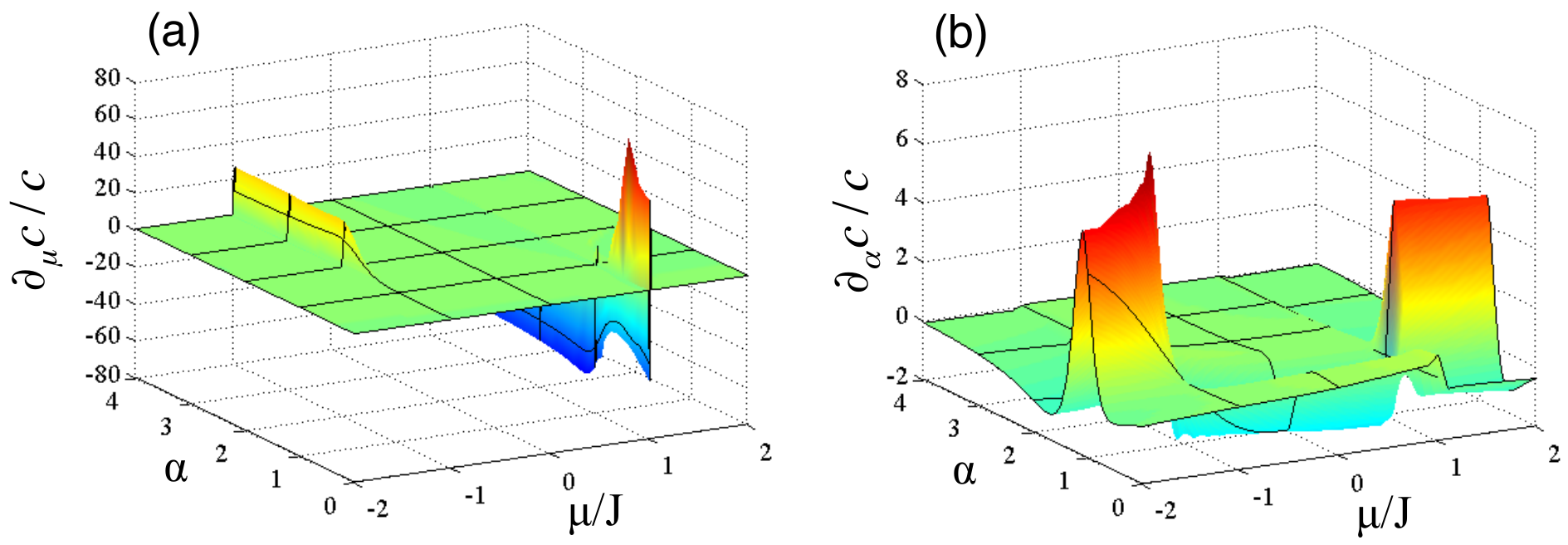}
\caption{Normalized derivative of $c$ with respect to $\mu$ (a) and $\alpha$ (b).}
\label{FigSupp4}
\end{figure}

\newpage

{\it Finite-size scaling analysis around $\alpha=1$.--}
From Fig.~\ref{FigSupp3} and the analysis of the scaling coefficients $b$ and the prefactor $c$, it is evident that the 
derivative of the Fisher density with respect to $\mu$ diverges at $\mu/J=1$ for $\alpha\geq 0$ and $\mu/J=-1$ for $\alpha>1$,
in correspondence of the QPTs.
It is less evident, from the numerical data, that the derivative of the Fisher density with respect to $\alpha$ diverges
at $\alpha=1$ and any value of $\mu$, in correspondence of the QPT.
We have thus performed a finite-size scaling analysis of the normalized derivative $\tfrac{1}{f_Q} \tfrac{d f_Q}{d \alpha}$ around $\alpha=1$ and different system size $L$.
From the results shown in Fig.~\ref{FigSupp12} we clearly see that, increasing $L$, slowly $\tfrac{1}{f_Q} \tfrac{d f_Q}{d \alpha}$ tends to peak at 
$\alpha=1$.

\begin{figure}[h!]
\includegraphics[clip,scale=1.1]{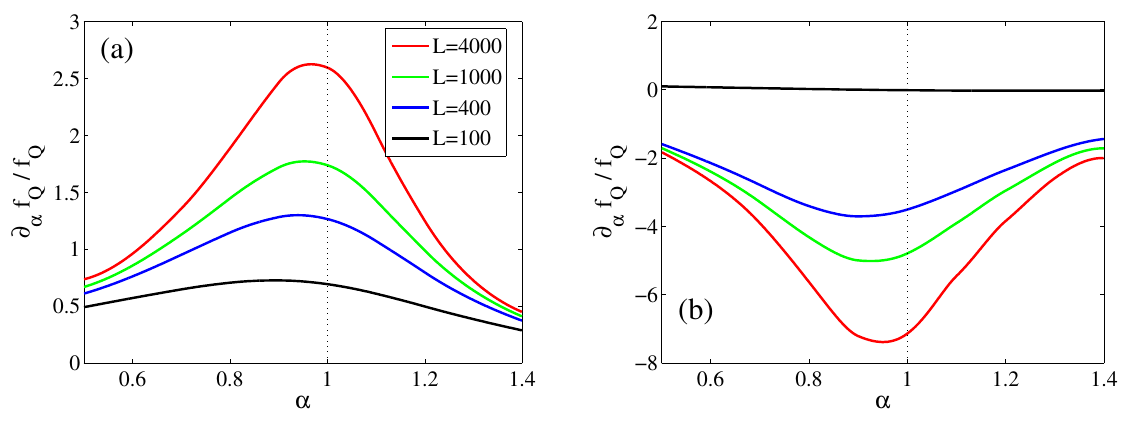}
\caption{Plot of $\tfrac{1}{f_Q} \tfrac{d f_Q}{d \alpha}$ as a function of $\alpha$ around the phase transition point $\alpha=1$ (vertical dotted line)
for different values of $L$.
Panel (a) is for $\mu/J=0$, panel (b) for $\mu/J=1.5$.
Increasing the system size, normalized derivative increases and tends to pick at $\alpha=1$.}
\label{FigSupp12}
\end{figure}

\newpage

{\it $\Delta/J$--$\mu/J$ phase diagram, $\alpha=\infty$.--}
In Fig.~\ref{FigSupp5} we show the Fisher density for the optimal operator in the $\Delta/J$--$\mu/J$ phase diagram
for short-range pairing ( $\alpha=\infty$).
In Fig.~\ref{FigSupp6} we show the corresponding coefficients $b$ and $c$.

\begin{figure}[h!]
\includegraphics[clip,scale=0.68]{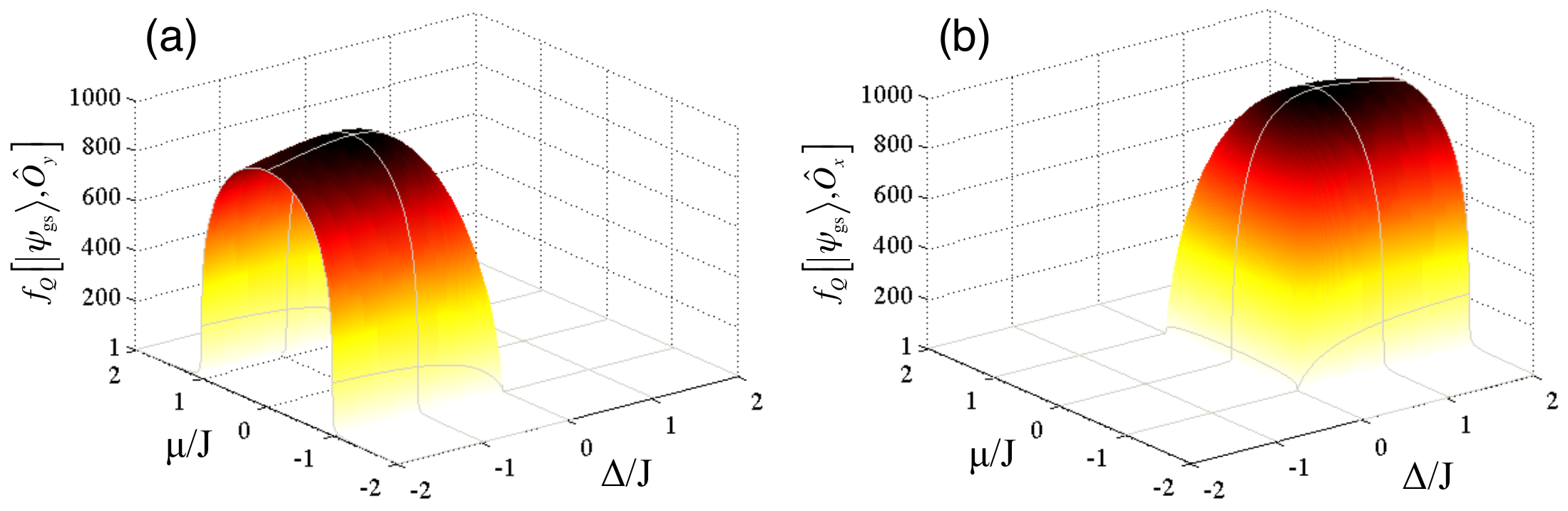}
\caption{(a) $f_Q[\vert \psi_{\rm gs} \rangle, \hat{O}_y]$ and (b)  $f_Q[\vert \psi_{\rm gs} \rangle, \hat{O}_x]$
in the $\Delta/J$--$\mu/J$ phase diagram, for $L=1000$ and $\Delta=J>0$.
Solid lines are cut at integer values of the parameters.}
\label{FigSupp5}
\end{figure}

\begin{figure}[h!]
\includegraphics[clip,scale=0.68]{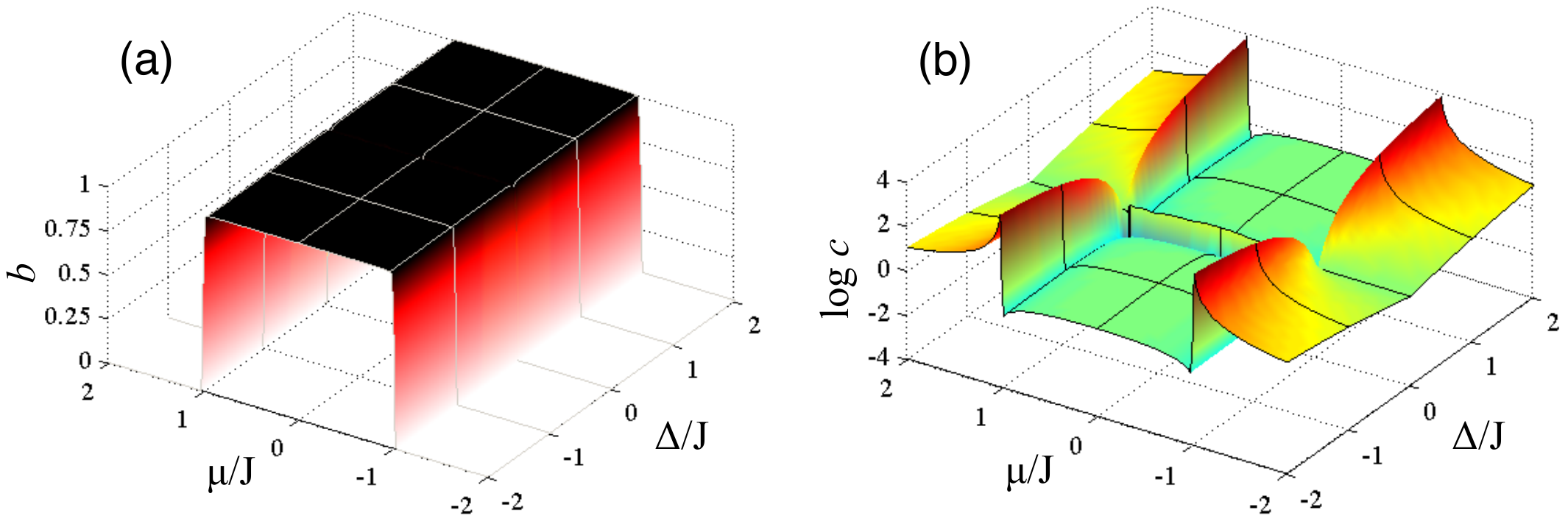}
\caption{Coefficient $b$ and $c$.
Solid lines are cut at integer values of the parameters.}
\label{FigSupp6}
\end{figure}

\newpage

{\it $\Delta/J$--$\mu/J$ phase diagram, $\alpha=0$.--}
In Fig.~\ref{FigSupp7} we show the Fisher density for the optimal operator in the $\Delta/J$--$\mu/J$ phase diagram
for long-range pairing ( $\alpha=0$).
In Fig.~\ref{FigSupp8} we show the corresponding coefficients $b$ and $c$.

\begin{figure}[h!]
\includegraphics[clip,scale=0.68]{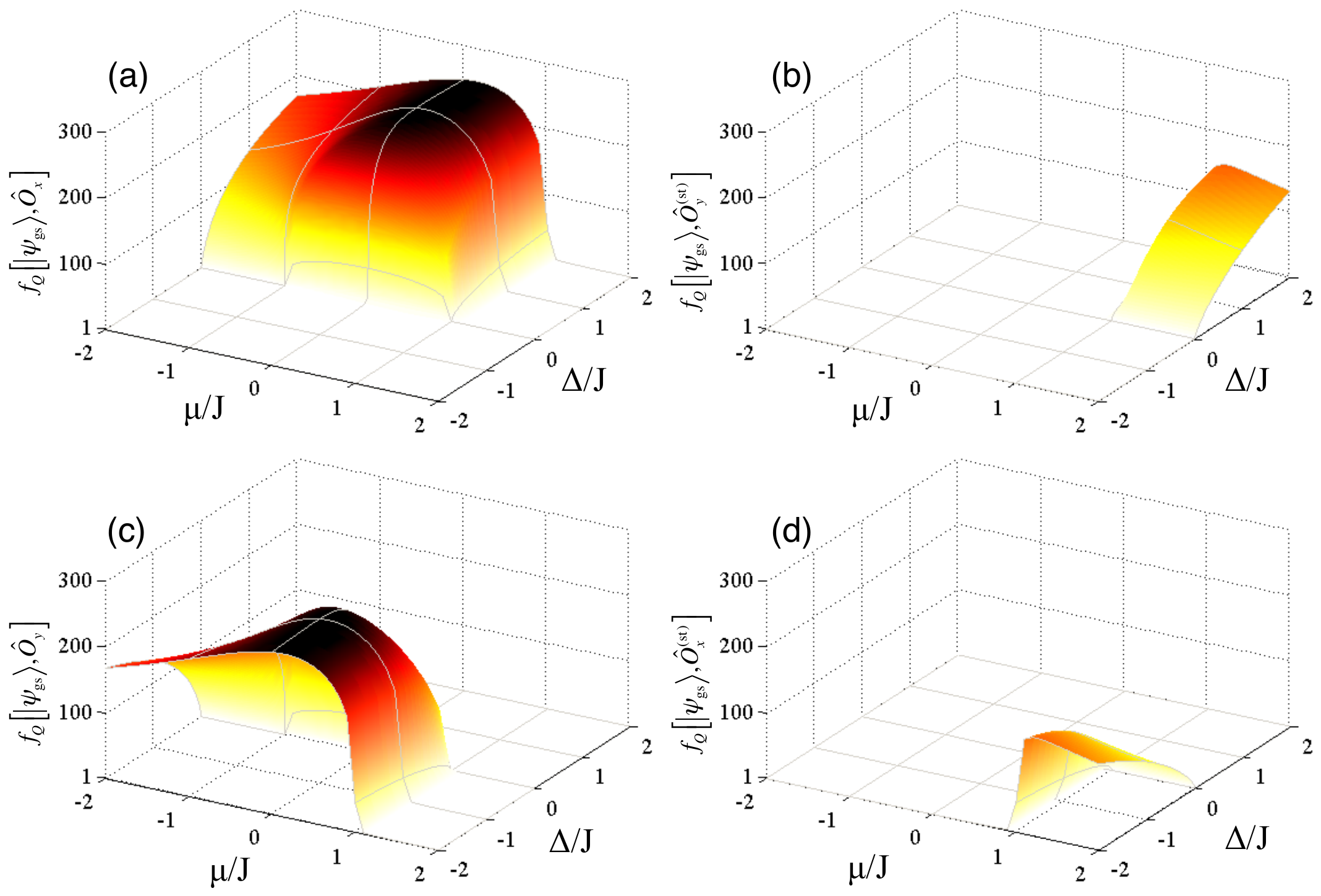}
\caption{(a) $f_Q[\vert \psi_{\rm gs} \rangle, \hat{O}_x]$ and (b)  $f_Q[\vert \psi_{\rm gs} \rangle, \hat{O}_y^{(\rm st)}]$
(c) $f_Q[\vert \psi_{\rm gs} \rangle, \hat{O}_y]$ and (d)  $f_Q[\vert \psi_{\rm gs} \rangle, \hat{O}_x^{(\rm st)}]$
in the $\Delta/J$--$\mu/J$ phase diagram, for $L=1000$ and $\Delta=J>0$.
Solid lines are cut at integer values of the parameters.}
\label{FigSupp7}
\end{figure}

\begin{figure}[h!]
\includegraphics[clip,scale=0.68]{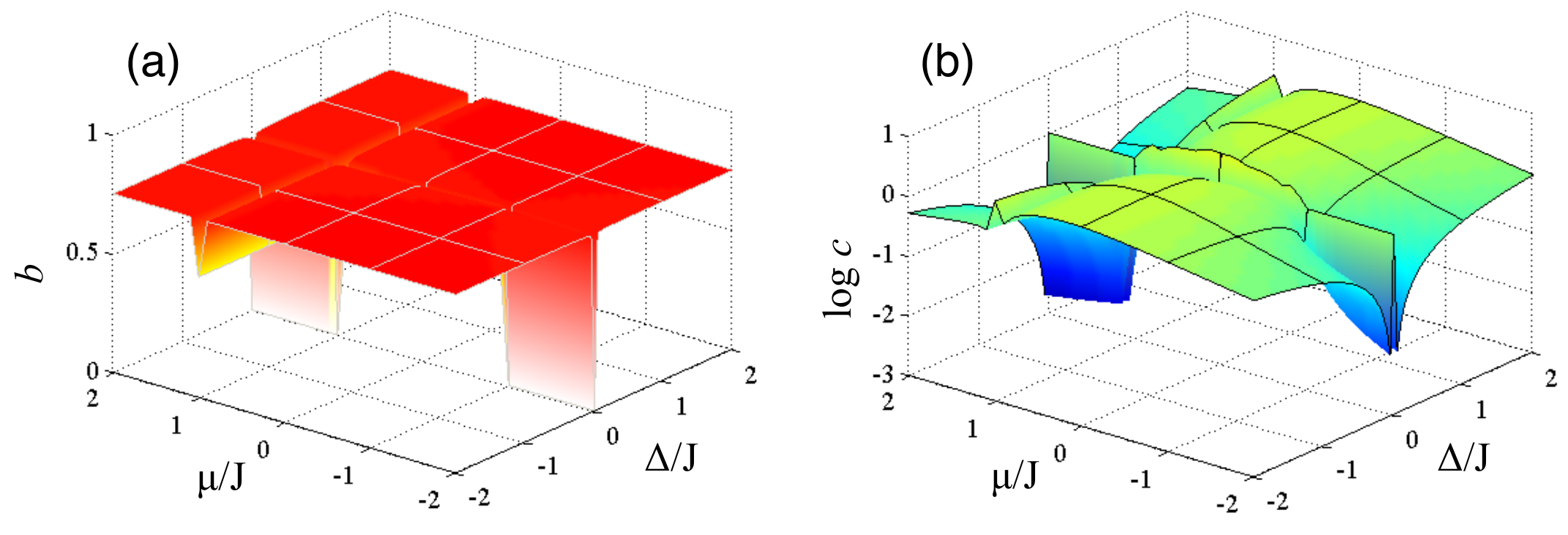}
\caption{Coefficient $b$ and $c$.
Solid lines are cut at integer values of the parameters.}
\label{FigSupp8}
\end{figure}

{\it Correlation functions and polynomial fits.--}
Below, we show polynomial fits of the Fisher density for different values of the parameters.
These and similar fits are used to extract the coefficients $b$ and $c$ discussed above and in the manuscript. 
We also show examples of the rescaling of the correlation function, highlighting the collapse of curves 
$L^{1-b} C_{\rho}(l)$ as a function of $l/L$ and for different values  of $L$.
This implies that the correlation function can be written as $C_{\rho}(l) = L^{b-1}c_{\rho}(l/L)$.
Computing Eq.~(5) gives
\be
f_Q[\vert \psi_{\rm gs}\rangle, \hat{O}_{\rho}] = 1 + L^{b-1} \sum_{l=1}^{L-1} c_{\rho}(l/L) =  1 + L^b \int_{0}^{1} dx \, c_{\rho}(x),
\ee
where the last equality is obtained in the limit $L\to \infty$.

\begin{figure}[h!]
\includegraphics[clip,scale=1.05]{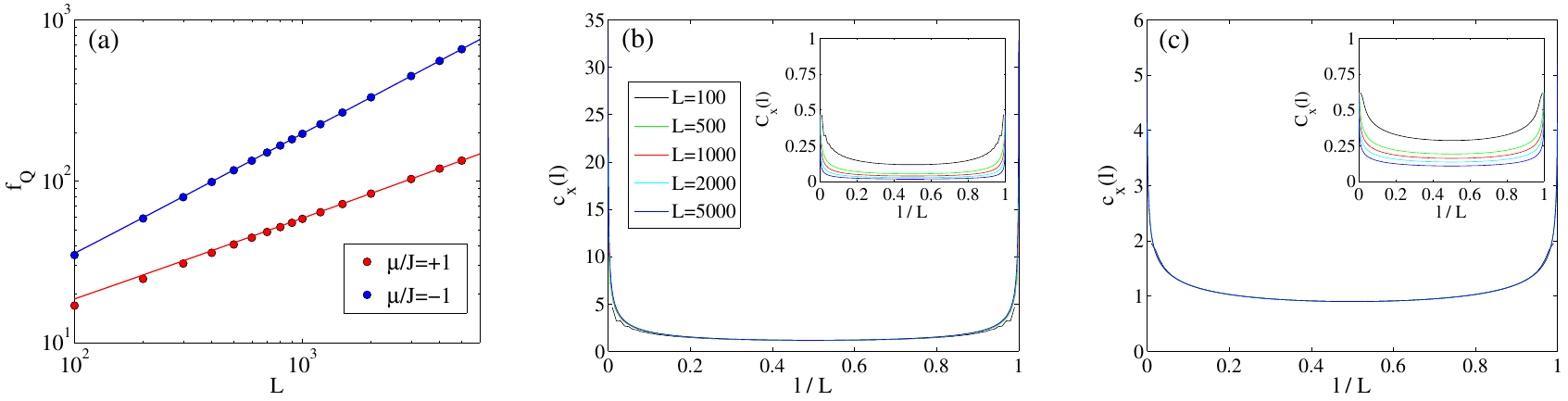}
\caption{Analysis for $\alpha=0.5$ and $\mu/J \leq 1$. 
Panel (a) shows numerical data of $f_Q[\vert \psi_{\rm gs} \rangle, \hat{O}_x]$ as a function of $L$ 
for $\mu/J=1$ (red dots) and $\mu/J=-1$ (blue dots).
The solid lines are polynomial fits to the data: $f_Q[\vert \psi_{\rm gs} \rangle, \hat{O}_x] = 1 + 1.63 \, L^{0.517}$ (red)
and $f_Q[\vert \psi_{\rm gs} \rangle, \hat{O}_x] = 1 + 1.09 \, L^{0.751}$ (blue).
Panels (b) and (c) show the correlation function $C_{x}(l)$ (inset) and the rescaled $c_{x}(l) = L^{1-b} C_{x}(l)$
(main) for different values of $L$ (see legend), $\mu=1$ (b) and $\mu=-1$ (c).}
\label{FigSupp9}
\end{figure}

\begin{figure}[h!]
\includegraphics[clip,scale=1.05]{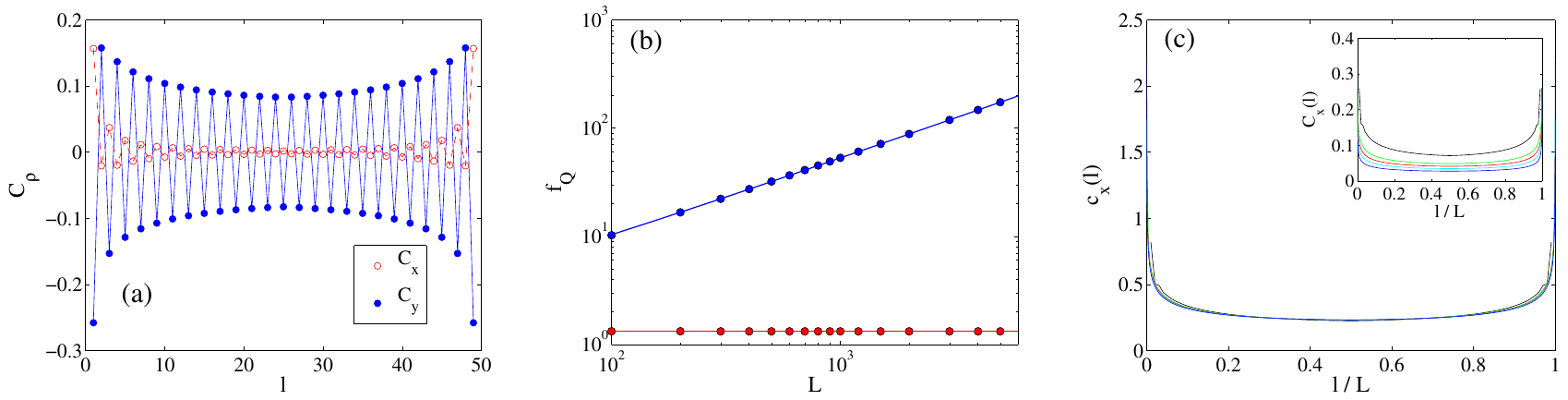}
\caption{Analysis for $\alpha=0.5$ and $\mu/J=2$. 
Panel (a) shows the correlation function $C_x(l)$ (red circles) and $C_y(l)$ (blue dots) for $L=50$.
Both functions are staggered but $C_y(l)$ is long-range. 
Panel (b) shows numerical data of $f_Q[\vert \psi_{\rm gs} \rangle, \hat{O}_x]$ (red dots) and  
$f_Q[\vert \psi_{\rm gs} \rangle, \hat{O}_y^{(\rm st)}]$ (blues dots) as a function of $L$.
The solid lines are polynomial fits to the data: $f_Q[\vert \psi_{\rm gs} \rangle, \hat{O}_x] = 1.323$ (red)
and $f_Q[\vert \psi_{\rm gs} \rangle, \hat{O}_y^{(\rm st)}] = 1 + 0.30 \, L^{0.745}$ (blue).
Panels (b) and (c) show the correlation function $(-1)^l C_{y}(l)$ (inset) and the rescaled $c_{y}(l) = L^{1-b} (-1)^l C_{y}(l)$
(main) for different values of $L$ [colors as in Fig.~{\ref{FigSupp9}}(b)].}
\label{FigSupp11}
\end{figure}

\begin{figure}[h!]
\includegraphics[clip,scale=1.05]{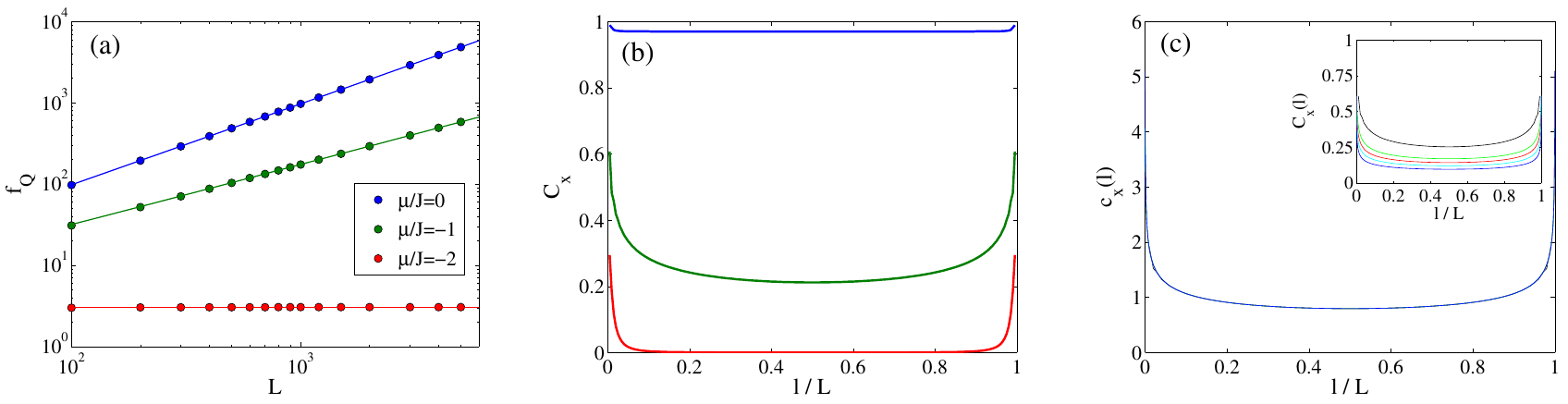}
\caption{Analysis for $\alpha=2$. 
Panel (a) shows numerical data of $f_Q[\vert \psi_{\rm gs} \rangle, \hat{O}_x]$ as a function of $L$ 
for $\mu/J=0$ (blue dots), $\mu/J=-1$ (green dots) and $\mu/J=-2$ (red dots).
The solid lines are polynomial fits to the data: $f_Q[\vert \psi_{\rm gs} \rangle, \hat{O}_x] = 1 + 0.96 \, L$ (blue),
$f_Q[\vert \psi_{\rm gs} \rangle, \hat{O}_x] = 1 + 0.96 \, L^{0.751}$ (green) and 
 $f_Q[\vert \psi_{\rm gs} \rangle, \hat{O}_x] = 3.04$ (red).
 Panel (b) shows the correlation function $C_{x}(l)$ for $L=200$, $\alpha=2$, $\mu/J=0$ (blue line), 
 $\mu/J=-1$ (green), and $\mu/J=-2$ (red line).  
 Panel (c) show the correlation function $C_{x}(l)$ (inset) and the rescaled $c_{x}(l) = L^{1-b} C_{x}(l)$
(main) for different values of $L$ (see legend), $\alpha=2$ and $\mu=-1$.}
\label{FigSupp10}
\end{figure}

\newpage

{\it Fidelity susceptibility.--}
The normalized ground state of the Kitaev chain can be written as
\be
\ket{\psi_{\rm gs}} = \prod_{n=0}^{L/2-1} \bigg( \cos \frac{\theta_{k_n}}{2} - \ii \sin \frac{\theta_{k_n}}{2} \hat{a}_{k_n}^{\dag} \hat{a}_{-k_n}^{\dag} \bigg) \ket{0},
\ee
where $\ket{0}$ is the vacuum of $\hat{a}_k$.
The ground state fidelity is 
\be
\mathcal{F} =  \bra{ \psi_{\rm gs} } \tilde{\psi}_{\rm gs} \rangle = 
\prod_{n=0}^{L/2-1}   \cos \bigg( \frac{\theta_{k_n} - \tilde{\theta}_{k_n}}{2} \bigg),
\ee
where $\ket{\psi_{\rm gs}} $ and $\ket{\tilde{\psi}_{\rm gs}}$ refer to the ground state for different values of the parameter(s).
Expanding in Taylor series  for $\tilde{\theta}_{k_n} \approx \theta_{k_n}$, up to second order in $d \theta_{k_n} = \tilde{\theta}_{k_n} - \theta_{k_n}$, we find
\be
\mathcal{F} \approx 1 - \frac{1}{8} \sum_{n=0}^{L/2-1} (d \theta_{k_n})^2.
\ee
Finally, the fidelity susceptibility with respect to the variation of a generic parameter $\eta$ is defined as 
\be
\chi_\eta = -  \frac{\ud^2 \mathcal{F}}{ \ud \eta^2} = \frac{1}{4} \sum_{n=0}^{L/2-1} \bigg(\frac{\ud  \theta_{k_n}}{\ud \eta}\bigg)^2,
\ee
where 
\be
\theta_{k_n} = \arccos \bigg( \frac{J \cos k_n + \mu}{\sqrt{ (J \cos k_n + \mu)^2 + \Delta^2 f^2_\alpha(k_n)/4 }} \bigg).
\ee
We now give the explicit expression of the fidelity susceptibility considering the variation of different parameters:
\begin{itemize}
\item varying $\mu$:
\be
\chi_\mu =  \frac{1}{4} \sum_{n=0}^{L/2-1}
\frac{\Delta^2 f^2_\alpha(k_n)/4}{[(J \cos k_n + \mu)^2 + \Delta^2 f^2_\alpha(k_n)/4 ]^2}.
\ee
\item varying $\alpha$:
\be
\chi_\alpha =  \frac{1}{4} \sum_{n=0}^{L/2-1}
\frac{(J \cos k_n + \mu)^2}{[(J \cos k_n + \mu)^2 + \Delta^2 f^2_\alpha(k_n)/4]^2 } 
\bigg( \frac{\Delta}{2} \frac{\ud f_\alpha(k_n) }{\ud \alpha} \bigg)^2
\ee
\item varying $\Delta$:
\be
\chi_\Delta =  \frac{1}{4} \sum_{n=0}^{L/2-1}
\frac{(J \cos k_n + \mu)^2}{[(J \cos k_n + \mu)^2 + \Delta^2 f^2_\alpha(k_n)/4]^2 }  \frac{f^2_\alpha(k_n)}{4}.
\ee

\end{itemize}

We want to see whether $\chi_\alpha$ diverges in the thermodynamic limit $L\to \infty$ when $\alpha \to 1$.
We first calculate $f_{\alpha}(k)$ in the limit $L\to \infty$, giving 
$f_{\alpha}^{\infty}(k) = [{\rm Li}_{\alpha}(e^{\ii k}) - {\rm Li}_{\alpha}(e^{-\ii k})]/\ii$, where ${\rm Li}_{\alpha}(x)$ are 
polylogarithmic functions. 
To see the divergence of $\chi_\alpha$ it is enough to see that one term in the sum diverges
(this because we have a sum of positive terms).
We calculate 
\be \label{chialphak}
\chi_\alpha (k) = \frac{(J \cos k + \mu)^2}{[(J \cos k + \mu)^2 + \Delta^2 f^{\infty}_\alpha(k)^2/4]^2 } 
\bigg( \frac{\Delta}{2} \frac{\ud f^{\infty}_\alpha(k) }{\ud \alpha} \bigg)^2
\ee 
in the limit $k = \pi/L \to 0$.
The results of calculations are shown in Fig.~\ref{FigSupp14}. 
We see that, very slowly in $L$,  $\chi_\alpha (k)$ peaks at $\alpha=1$.
Exploiting the expansion $f_{\alpha}^{\infty}(k) \sim k^{\alpha-1}$ for 
$ k \to 0$, it is easy to check that the found peak in  $\chi_\alpha (k)$, developing as $(\log L)^2$, is directly related with the divergence of $f_{\alpha}^{\infty}(k)$ at $\alpha = 1$;
the same divergence, holding also at $\alpha<1$, has been found also responsible for the appearance of the semi-integer winding numbers in the same regime.

This result for the divergence of $\chi_\alpha (k)$ justifies the emergence of new phases below $\alpha =1$, even if no vanishing mass gap in the Bogoliubov spectrum occurs at this threshold. 

\begin{figure}[h!]
\includegraphics[clip,scale=1]{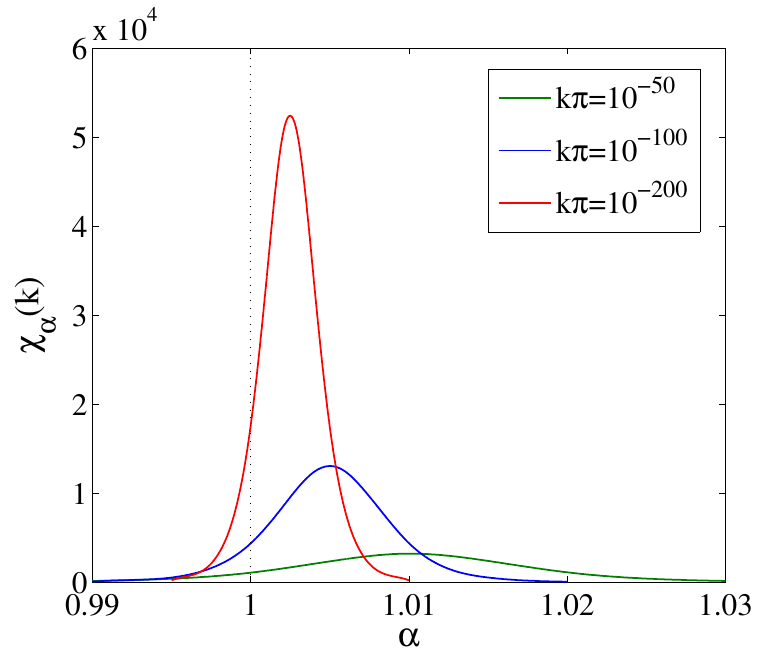}
\caption{Plot of Eq.~(\ref{chialphak}) around $\alpha=1$ (vertical dotted line). 
Different solid lines refer to different values of $k$ in the limit $k \to 0$.
Here $\Delta=J$ and $\mu/J=0$. Similar results can be found for other values of $\mu$.}
\label{FigSupp14}
\end{figure}

\end{widetext}


\begin{thebibliography}{400}
\bibitem{footnote3}
In the case $\Delta<0$ the phase diagram is obtained changing $W \to -W$.
\end{thebibliography}
\end{document}